\documentstyle[multicol,aps,epsf,here]{revtex}
\tighten

\begin{document}

\newcommand{\half}{\frac {1}{2} }
\newcommand{\eg}{{\em e.g.} }
\newcommand{\ie}{{\em i.e.} }
\newcommand{\etc}{{\em etc.}}
\newcommand{\si}{\simeq}
\newcommand{\etal}{{\em et al.\ }}
\newcommand{\cf}{{\em cf. }}

\newcommand{\dd}[2]{{\rmd{#1}\over\rmd{#2}}}
\newcommand{\pdd}[2]{{\partial{#1}\over\partial{#2}}}
\newcommand{\pa}[1]{\partial_{#1}}
\newcommand{\pref}[1]{(\ref{#1})}

\newcommand {\be}[1]{
%%      For draft    %%%%%%%%
%{\marginpar{{\scriptsize\ \\ \ #1}}}
      \begin{eqnarray} \mbox{$\label{#1}$}}

\newcommand{\ee}{\end{eqnarray}}

\def\rdown{\rho_{\downarrow}}
\def\pa{\partial}
\def\th{\theta}
\def\bea{\begin{eqnarray}}
\def\eea{\end{eqnarray}}
\def\be{\begin{equation}}
\def\ee{\end{equation}}
\def\pa{\partial}
\def\eps{\epsilon}
\def\th{\theta}
\def\na{\nabla}
\def\nn{\nonumber}
\def\lan{\langle}
\def\ran{\rangle}
\def\pr{\prime}
\def\rarrow{\rightarrow} 
\def\larrow{\leftarrow} 

%%%%%%%%%%%%%%%%%%%%%
%\vspace* {-21 mm}
%\begin{flushright}
%  NORDITA-2000/XX CM \\
%  April-2000 \\
%\end{flushright}
%%%%%%%%%%%%%%%%%%%%

\title{Electron correlation effects in the presence of non-symmetry
dictated nodes}
\vskip 20 mm
\author{P. Singha Deo} 
\address{S.N. Bose National Centre for Basic Sciences,
         J.D.Block, Sector III, Salt Lake City, Calcutta 700098, India} 
\date{\today}

\maketitle

\begin{abstract} 

We numerically study the effect of non-symmetry-dictated-nodes (NSDN)  on
electron correlation effects for spinless electrons. We find that
repulsive interaction between electrons can enhance the overlap between
nearest neighbors in the tight binding Hamiltonian, in the presence of
NSDN. Normally, in the absence of NSDN, attractive interaction between
electrons give such an effect and repulsive interaction gives opposite
effect.

\hfill\\ 
%PACS: 73.23.-b, 72.10.-d, 72.10.Bg
\end{abstract}
\begin{multicols}{2}

\narrowtext
%%%%%%%%%%%%%%%%%%%%%%
%%%%%%%%%%%%%%%%%%%%%%

At low temperatures, inelastic collisions are greatly suppressed.  As a
result the phase coherence length of an electron can become a few microns. 
Mesoscopic systems are defined as systems in which the phase coherence
length exceeds the sample size. In such a system elastic scattering is the
dominant feature and such mesoscopic samples can be understood as phase
coherent elastic scatterers.  Recently a new kind of scattering phase
shift was discussed in Q1D, in connection with the violation of the parity
effect \cite{deo96}. To explain this phase shift and the violation of
parity effect, we first explain what are symmetry dictated nodes (SDN) and
non-symmetry dictated nodes (NSDN) that can arise in a Q1D system. 

For example let us consider a rectangular quantum billiard or dot
connected to leads by quantum mechanical tunneling as shown in Fig.1.  The
system has reflection symmetry across the x-axis as well as the y-axis. 
Also the x-y components separate and spanning nodes (nodes that span
across the direction of propagation and shown by dashed line in Fig.1) as
well as non-spanning nodes (shown by dotted line in Fig.1)  develop in the
geometry, dictated by the reflection symmetries. There are various
symmetries that give rise to nodes in the wave-function and we call them
SDN (for example the antisymmetric property of the many body wave function
result in nodes). But if quantum coherence extends to some distance inside
the leads then one can model the phase coherent quantum dot as shown in
Fig.2.  Note that in this case also reflection-symmetry holds in the
x-direction as well as in the y-direction, but x-y components do not
separate. So by tuning the boundary condition in y-direction by a gate
voltage one can develop nodes that try to develop across y-direction but
also act across x-direction and change the phase of the wave function in
x-direction by $\pi$.  We call them NSDN because they do not originate due
to the symmetry of the Hamiltonian but originates due to the boundary
conditions. 

There are many configurations of such NSDN, the simplest one was discussed
for the stub geometry (shown in Fig.3) in Ref. \cite{deo96}, and further
elaborated here.  Consider for example two finite one dimensional quantum
wires of equal length, AB and CD placed perpendicular to each other as
shown in Fig.3 by the solid lines.  When CD is completely detached from AB
as shown in Fig.3(a), then the quantum mechanical wave function in AB and
CD in the ground state is shown by the dotted lines. They are basically
the ground state wave function in an infinite potential well in one
dimension (1D). As is known to us, the ground state wave functions are by
symmetry, even parity states without any nodes, except at the boundary. 
But when CD is attached to AB to give a T-shaped stub structure as shown
in Fig.3(b), then CD forms a node at C, which is also the midpoint of AB. 
The wave function in this case is again shown by dotted lines and the wave
function between A and B is no longer an even parity state but an odd
parity state. The node at C between A and B does not originate from the
symmetry of the Hamiltonian and is not a symmetry dictated node (SDN). It
is rather forced by the boundary condition in the y-direction and is a
NSDN. An infinitesimal change in the length CD makes this node disappear
and then we have no node between A and B.  The node at C induces a phase
change by $\pi$ and when we join A and B together to form a ring-stub
system, we also get persistent currents without parity effect, as parity
of the persistent currents is sensitive to the number of nodes in the wave
function \cite{leg91}. 

When semi infinite leads are attached to A and B in Fig.3(b)  then the
transmission amplitude of the stub structure has zero-pole pair and one
gets Fano resonances \cite{tek93}.  Fano resonances \cite{fan} are a very
general feature of Q1D \cite{bag90,deo98} systems in the presence of
defects and the Fano resonances are characterized by a zero-pole pair of
the transmission amplitude in the complex energy plane.  At the energy
corresponding to the pole a charge gets trapped by the scatterer and there
is also an energy when there is a zero transmission across the scatterer.
At the energy corresponding to the zero, scattering phase shift
discontinuously changes by $\pi$ due to the NSDN \cite{deo96}.  It seems
at present that this phase due to NSDN is necessary to understand the
experimental results of Refs. \cite{yac95,sch97,CER97}. Initial analysis
of the experiments in terms of the captured charge and Friedel-sum-rule
\cite{yey95} revealed the shortcomings of applying Friedel-sum-rule to the
quantum dot.  The phase change due to NSDN can explain the experimental
data, was first proposed in Ref. \cite{jay96} and later discussed in Refs. 
\cite{deo98,cho98,xu98,lee99,tan99,yeycm}. Friedel sum rule in the
presence of this NSDN has also been generalized in Refs. 
\cite{lee99,tan99,swa01}. 

In this work we intend to show that this NSDN greatly affects electron
correlation effects. We will concentrate on the situation when we have
NSDN in the ring-stub geometry, that is when points A and B in Fig. 3(b) 
are joined together to form a ring ACB coupled to a stub CD. In order to
diagonalize the electronic Hamiltonian (unlike Ref. \cite{sta} we are not
interested in the positive charges and polarization) we will consider a
lattice Hamiltonian as given below where the ring is made up of three
sites (located at A, B and C) and the stub is made up of two sites
(located at E and D)  (see the inset to Fig.~4).  Such a choice allows a
NSDN exactly at the Fermi surface as we shall soon see. The isolated site
energies are 0 and ignoring electron spin the Hamiltonian is \be H=-t
\Sigma_{<ij>} c_i^+c_j + u \Sigma_{<ij>} n_in_j \ee where $c_i$ is the
electron destruction operator at site $i$, $n_i=c_i^+c_i$ is the operator
that gives the occupation probability of site $i$ and $<ij>$ means only
nearest neighbors are considered in the sum.  The two competing energy
scales of an interacting electron system are parametrized by $t$ and $u$. 
The possible way in which these parameters $t$ and $u$ modify electron
correlation effects is known. For positive values, $u$ plays the role of
the coulomb interaction strength between two electrons, that tend to make
the electrons avoid each other and localize them as far from each other as
possible while $t$ plays the role of the quantum mechanical kinetic energy
of the electrons that tends to make the electrons overlap with each other
and delocalize. We choose the unit of energy such that $t=1$, through out
our calculations.

In Fig.~4 we plot the single particle energies of the system when $u/t$=0,
as a function of a flux that penetrates through the centre of the ring
ABC. It is known that the single particle states will disperse with the
flux with a periodicity of $\alpha=2 \pi \phi / \phi_0 =2 \pi$ \cite{bil}. 
Here $\phi$ is the flux through the ring and $\phi_0=hc/e$, is the flux
quantum.  This is why we plot the single particle eigen energies, which
will be five in number (shown by five solid curves)  as there are five
sites in all, from $\alpha=-\pi$ to $\alpha=\pi$ (\ie in the first
Brillouin zone).  The ground state is diamagnetic as expected, \ie
increases in energy with the flux.  Striking feature of the spectrum is
that two of the states (the 1st excited state and 3rd excited state)  are
independent of the flux. This is a purely quantum effect and analyzed in
Ref. \cite{sre}. Wavefunctions in these states have NSDN, similar to what
is shown in Fig. 3 (b), except that now A and B are joined. Since the node
at C is a NSDN, the wave function does not extend across it and so does
not enclose the flux and do not depend on the flux. This does not happen
for states with SDN.  Hence if we put two electrons or four electrons in
the system then the NSDN will lie exactly at the Fermi energy.  While in a
finite system with discrete levels and large separation between single
particle levels, one can tune the fermi energy to coincide with the NSDN
only for some special parameter values, it will not be so in a large
system or in a system connected to leads.

If we put three electrons in the system then the Fermi surface will be far
away from the NSDN and when we switch on $u$ we will expect a competition
between $t$ and $u$. Although the outcome of this competition in the
absence of NSDN, is well known we will demonstrate it here in order to
accentuate how the NSDN can change this outcome.  For very large positive
$u$ one electron stays in the stub and two electrons stay in the ring,
there being very little overlap between the stub and the ring. This agree
with Ref. \cite{sta}, although the cases that can be studied within the
model there corresponds to the two electron case and not the three
electron case as considered here.  For example, in Fig.~5 we plot $<n_i>$
at sites D and E versus $\alpha$ for different values of $u>0$.  The
dotted curve gives $<n_D>$, \ie, expectation value of the operator $n_i$
at the site D when $u/t$=0.  The short-dashed curve gives $<n_E>$, \ie,
expectation value of the operator $n_i$ at the site E when $u/t$=0.  The
solid curve gives $<n_D>$, when $u/t$=1.  The dashed curve gives $<n_E>$,
when $u/t$=1.  The solid curve is greater than the dotted curve at all
flux values, i.e., $<n_D>$ is enhanced by $u$, while the dashed curve is
less than the short-dashed curve for all flux values, \ie $<n_E>$ is
reduced by $u$. Also note that the enhancement in $<n_D>$ is roughly the
same as the reduction in $<n_E>$ \ie enhancement in $<n_D>$ is at the cost
of $<n_E>$ and transfer of charge between the stub and the ring is
negligible.  This means repulsive interaction kills the overlap of the
wave-function between the sites D and E. For the known competition between
$t$ and $u$, the single electron in the stub is localized and resides
almost entirely in the site D.

Opposite effect will be seen for attractive interaction or for negative
$u$. For large attractive interaction all the three electrons will try to
cluster together and since the stub is made up of two sites only and
cannot accomodate three electrons, for large negative $u$ all the three
electrons will stay in the ring.  Hence for large and negative $u$, charge
transfer between the ring and the stub becomes dominant and hides the
effect that we want to study.  But for small negative $u$ \ie $|u|/t<1$,
interaction will try to enhance the occupation probability at nearest
neighbors.  For example, in Fig.~6 we plot $<n_i>$ at sites D and E versus
$\alpha$ for different values of $u<0$.  The dotted curve gives $<n_D>$,
when $u/t$=0.  The short-dashed curve gives $<n_E>$, when $u/t$=0.  The
solid curve gives $<n_D>$, when $u/t$=-0.5.  The dashed curve gives
$<n_E>$, when $u/t$=-0.5.  The solid curve is less than the dotted curve
at all flux values while the dashed curve is greater than the short-dashed
curve for all flux values.  So for small values of $u<0$ one can clearly
see the fact that $u$ tends to make $<n_E>$ to be roughly equal to
$<n_D>$, with small deviations due to charge transfer between the ring and
the stub, \ie $<n_D>+<n_E>$ is not conserved.

Next we consider the two electron system, \ie the situation when the NSDN
lies at the Fermi surface. In Fig~7 we plot $<n_{stub}>$=$<n_D>+<n_E>$
when $u/t$=0 (solid curve), $u/t$=1 (dotted curve) and $u/t$=100 (dashed
curve). Note that in the $u/t=0$ case for $|\alpha|<|\alpha^*|$,
$<n_{stub}>$ is roughly constant in $\alpha$ and then it drops, signifying
charge transfer between the stub and the ring. This is the case studied in
Ref. \cite{sta}, however, unlike that in Ref. \cite{sta}, our calculations
are exact and we consider the overlap between all possible states. When
the stub and the ring are decoupled from each other, then the states in
the ring will depend on $\alpha$ but the states of the stub will not.  So
as a ring state change in energy with $\alpha$ it can cross a stub state. 
In case of the parameters used in Fig.~7, for $|\alpha|<|\alpha^*|$, the
isolated stub ground-state is lower in energy than the isolated ring
first-excited-state while it is the opposite for $|\alpha|>|\alpha^*|$. If
the ring and the stub were decoupled, for $|\alpha|<|\alpha^*|$, the total
energy is minimized when we put one electron in the ground state of the
isolated stub and one electron in the ground state of the isolated ring. 
But for $|\alpha|>|\alpha^*|$, total energy is minimized if we put both
the electrons in the isolated ring, one in the ground state and the other
in the first excited state, the isolated stub being empty.  This is why
the drop in $<n_{stub}>$ occurs in the coupled ring-stub system at
$|\alpha|$=$|\alpha^*|$, where of course the kinetic energy parametrized
by $t$ tends to make the isolated ring and isolated stub states overlap,
the overlap being the maximum when the isolated ring state and the
isolated stub state become degenerate at $|\alpha^*|$. Also the NSDN
results as a consequence of this degeneracy between the isolated ring
state and the isolated stub state and affects the energetics of the system
specially around $|\alpha^*|$. Now if we switch on $u$, then as discussed
before, it will try to destroy the overlap between the ring and the stub
states.  We can see from Fig.~7 that when we increase $u$ then
$|\alpha^*|$ is pushed more towards the zone boundary and for very large
$u$ once again one electron is localized in the stub and one electron in
the ring as if the stub and the ring are decoupled.  One would expect that
$u$ will also destroy the overlap between the sites E and D and the
electron in the stub will reside mostly in one of the two sites as $u
\rightarrow \infty$.  We will show below that actually the opposite
happens due to the presence of the NSDN. 

In Fig.~8, we plot $<n_D>$ (dotted curve) and $<n_E>$ (short-dashed
curve), when $u/t=0$, for the two electron system. We then consider a
situation when $u$ is small and charge transfer between the ring and the
stub is not important.  When $u/t=1$, then $<n_D>$ is given by the thin
solid curve while $<n_E>$ is given by the long-dashed curve. One can see
that $u$ has decreased $<n_D>$ and enhanced $<n_E>$ and trying to
distribute the electron evenly between the two sites. That is a repulsive
interaction between nearest neighbors is enhancing the overlap rather than
destroying the overlap between these two nearest neighbor sites.  One can
actually see that the solid curve and the dashed curve are equal to each
other around $|\alpha^*|$, \ie, when the single particle states of the
isolated ring and the isolated stub are degenerate to each other and
result in a NSDN at that energy. Now let us consider a situation when $u$
is very large and hence the stub and the ring has one electron each and
there is absolutely no charge transfer between the stub and the ring. When
$u/t=100$, then $<n_D>$ is given by the dash-dotted curve, while $<n_E>$
is given by the dot-dot-dashed curve. For infinite $u$, when the NSDN is
pushed to the extreme end of the zone boundary, \ie $|\alpha^* |=\pi$, and
one electron is completely localized in the stub, then at $|\alpha^*
|=\pi$, once again $<n_D>$=$<n_E>$. For such large values of $u$, at small
$\alpha$ it can be seen that the electron in the stub mostly reside at the
site D as expected but not at $\alpha=\alpha^*$. 

The reason is as follows. The NSDN costs a lot of energy, and if the
system can get rid of this NSDN then it gains in energy. When interaction
is increased, then the stub electron has two options. Firstly, it can
mostly reside in the the site D in order to reduce the interaction energy,
but this does not help it to remove the NSDN. The NSDN merely moves into
the stub and resides at site E rather than the site C. The other option is
that the electron in the stub resides equally in the sites E and D, and
the NSDN is pushed above the Fermi energy and does not affect the
energetics of the system.  The interacting electrons can be now thought to
be as moving in an effective potential, and the single particle states of
the effective potential are like the states of an isolated stub and an
isolated ring, with one electron in each of them. But the effective
potential is such the electron in the ring and the electron in the stub
are nowhere degenerate. The degeneracy is removed by the way in which $u$
renormalized the real potential to give the effective potential. This
renormalization, populates the two stub sites equally and raises the
energy of the electron localized in the stub with respect to the electron
localized in the ring, such that the stub electron and the ring electron
are not degenerate at any flux.  In the absence of the degeneracy, the
NSDN also do not exist or rather becomes marginal as in this case.  The
NSDN can be pushed to the extreme zone boundary but cannot be pushed
beyond that for the presence of a gap in the spectrum. Hence however large
$u$ we choose we see some effect of the NSDN. But in a system with a
continuous spectrum it would be possible to remove the NSDN completely.

Thus we find that repulsive interaction in presence of NSDN can result in
some features that are typically like that of attractive interaction
between electrons. In this respect one may recall the fact that in a BCS
superconductor also electrons can exhibit an attractive interaction
between each other. They interact via an effective potential that is
actually negative. Whereas in the present case studied by us, there is no
effective potential that is negative. The interacting system always has a
free energy higher than that of the non-interacting system, in the
presence as well as in the absence of the NSDN.  However, in the presence
of the NSDN, the qualitative behavior of electron correlation effects is
like that of an attractive potential. It is seen in high Tc
superconductors, that electrons form pairs while there is no trace of any
attraction between them. So probably one should study the possibilities of
NSDN in two dimensional lattices.

The system considered here (i.e., three sites in the ring, two sites in
the stub and 2 electrons in the whole system)  gives a very good picture
of the competition between repulsive interaction between electrons and
NSDN. In larger systems, where the electrons tend to have greater degrees
of freedom, the charge transfer between the ring and the stub often hides
the effect. Numerical calculation suggest that the effect is still present
but it is often difficult to draw a clear conclusion by isolating the
effect.  Of course, there are infinite possibilities of choosing the
system parameters and it may be possible to find some regimes where charge
transfer is negligible and at the same time it is possible to tune the
NSDN to lie exactly at the Fermi surface. Also, we have not considered the
electron spin in the present study. The conclusions of an effective
attraction between repulsive electrons of the same spin can be drawn on
the basis of the study here. This is essentially because inclusion of spin
will not destroy the effective attraction between electrons with the same
spin. Of course there may be additional effects of NSDN-created effective
attraction or repulsion between electrons with opposite spins. We are
looking at such possibilities at present. Besides, for large values of
$u/t$, the stub can be occupied by only one electron with a given spin, a
phenomenon known as Coulomb Blockade and we believe the effect mentioned
here can be experimentally verified. One should look for polarization
effects of relative parts within the stub connected to a ring. The focus
of Ref.  \cite{sta} was to look for the polarization of the stub with
respect to the ring. Or one can also take two quantum dots, one at the
position of site E and connected to the ring on one side and to another
dot on the other side, the second dot being located at the site D. In such
a geometry, one can study the polarization of the two dots with respect to
each other, which will show the effects of this NSDN induced effective
attraction. Such a study will be reported in the future.

\end{multicols}

{\bf FIGURE CAPTIONS:}

\noindent {\bf Fig.~1.} A rectabgular quantum billiard or quantum dot,
weakly coupled to leads. The dotted line is along the x-axis
and the dashed line is along the y-axis.\\
{\bf Fig.~2.} A more realistic model for the quantum dot of Fig.~1.\\
{\bf Fig.~3.} Two one dimensional quantum wires of equal lengths,
AB and CD, shown by solid lines, placed along x and y directions,
respectively. The origin (x=0, y=0) is at the mid point of AB.
(a) CD is not connected to AB. (b) CD is connected to the mid
point of AB.\\
{\bf Fig.~4.} The inset shows a ring-stub geometry. When the ring
is made up of 3 sites and the stub is made up of 2 sites,
then there will be 5 single particle states in all. The energies ($E$)
of these five states as a function of a dimensionless
flux $\alpha$, through the centre
of the ring is shown by the five solid curves.\\
{\bf Fig.~5.} This figure shows the expectation values of the operator
for occupation probability at the sites E and D as a function
of $\alpha$. Here $u/t \ge 0$ and there are three electrons
in the system.\\
{\bf Fig.~6.} This figure shows the expectation values of the operator
for occupation probability at the sites E and D as a function
of $\alpha$. Here $u/t \le 0$ and there are three electrons
in the system.\\
{\bf Fig.~7.} This figure shows the sum of the
expectation values of the operator
for occupation probability at the sites E and D as a function
of $\alpha$. Here $u/t \ge 0$ and there are two electrons in the
system.\\
{\bf Fig.~8.} This figure shows the expectation values of the operator
for occupation probability at the sites E and D as a function
of $\alpha$. Here $u/t \ge 0$ and there are two electrons
in the system.\\
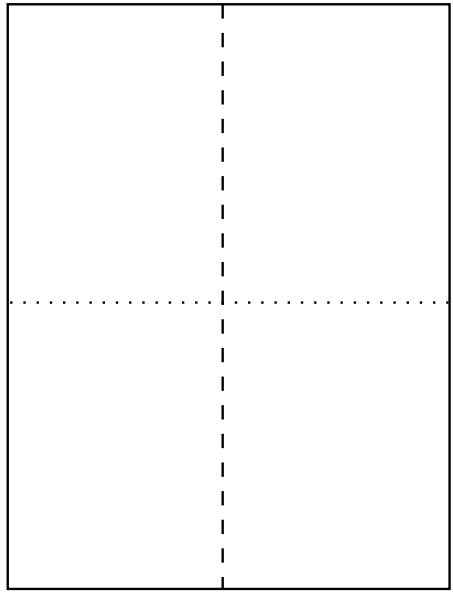
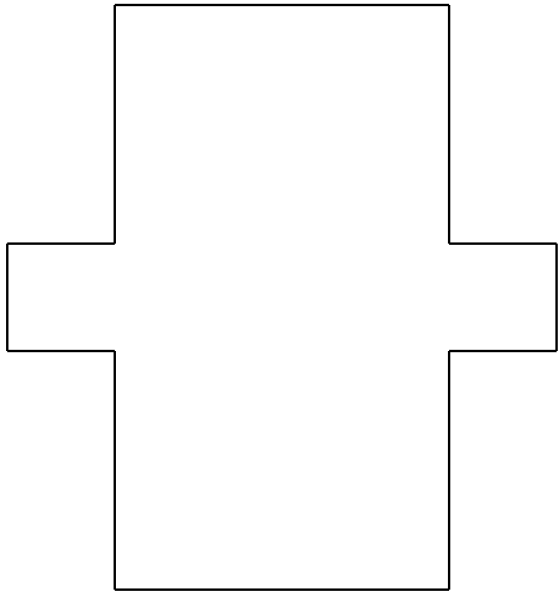
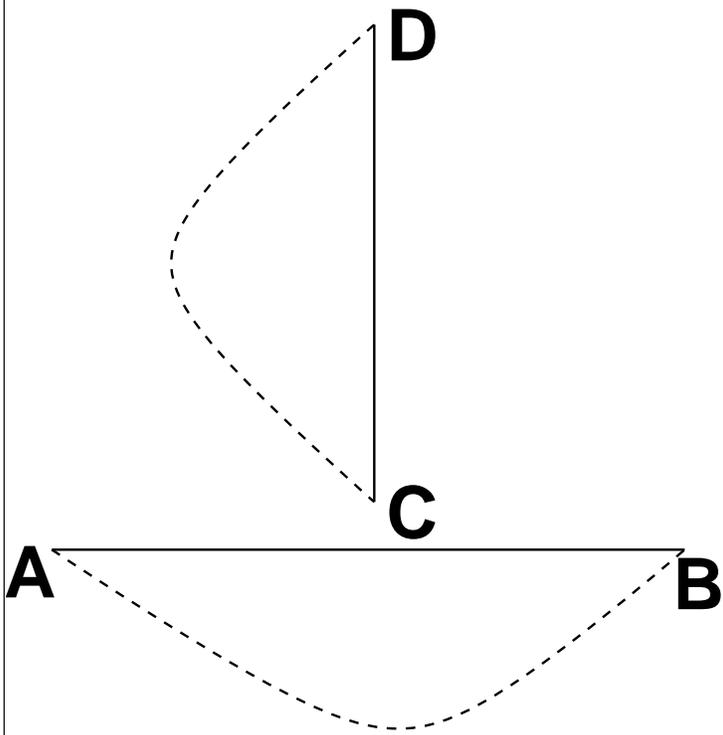
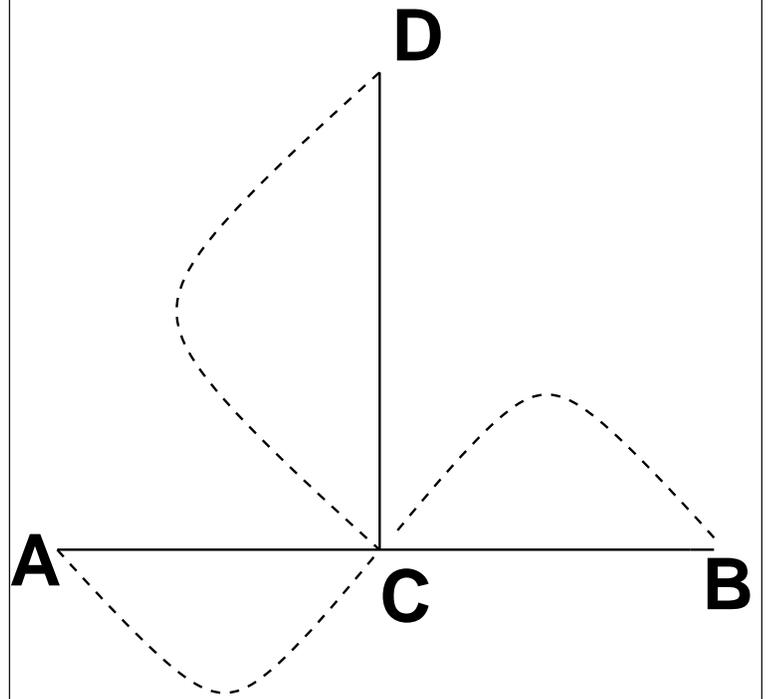
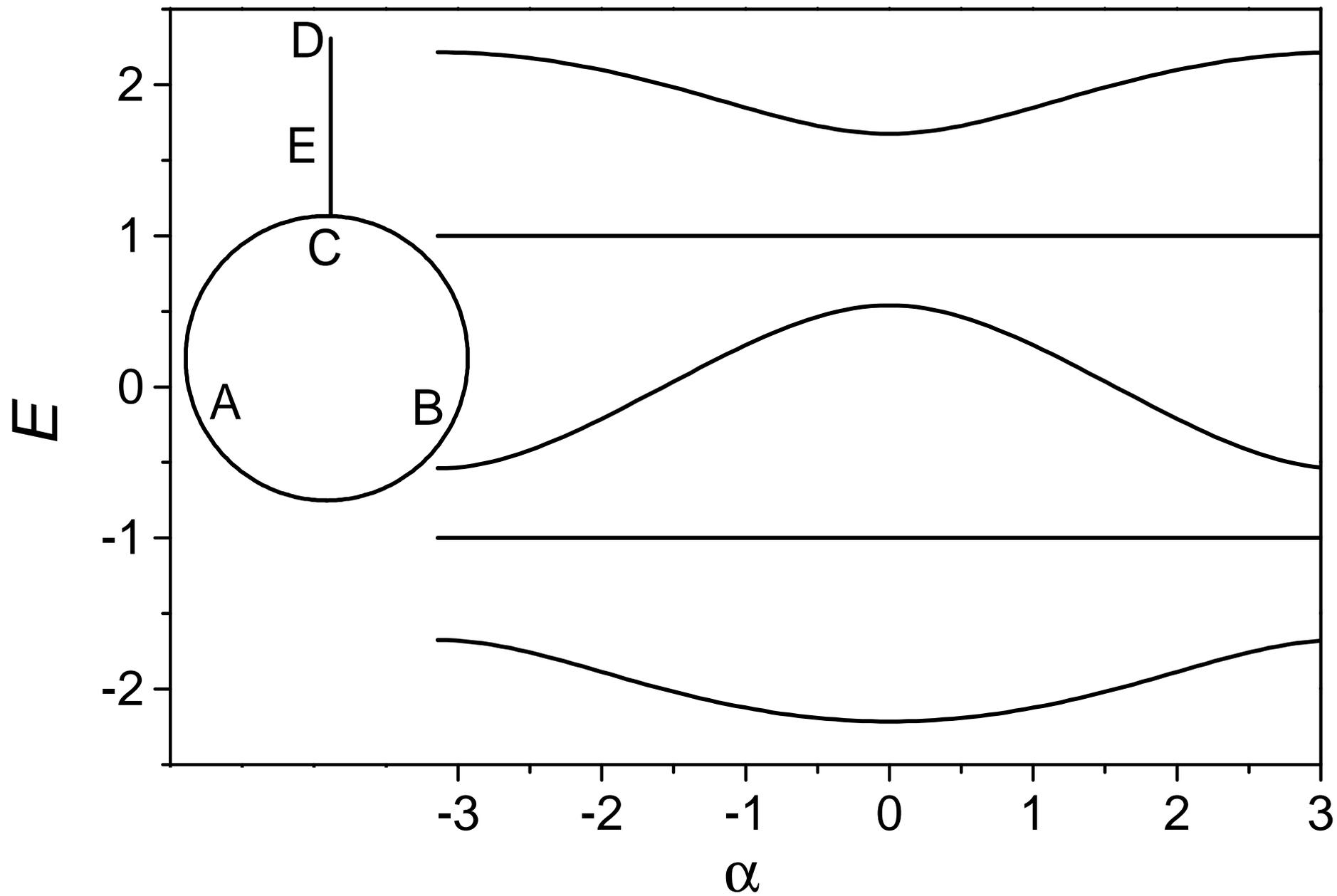
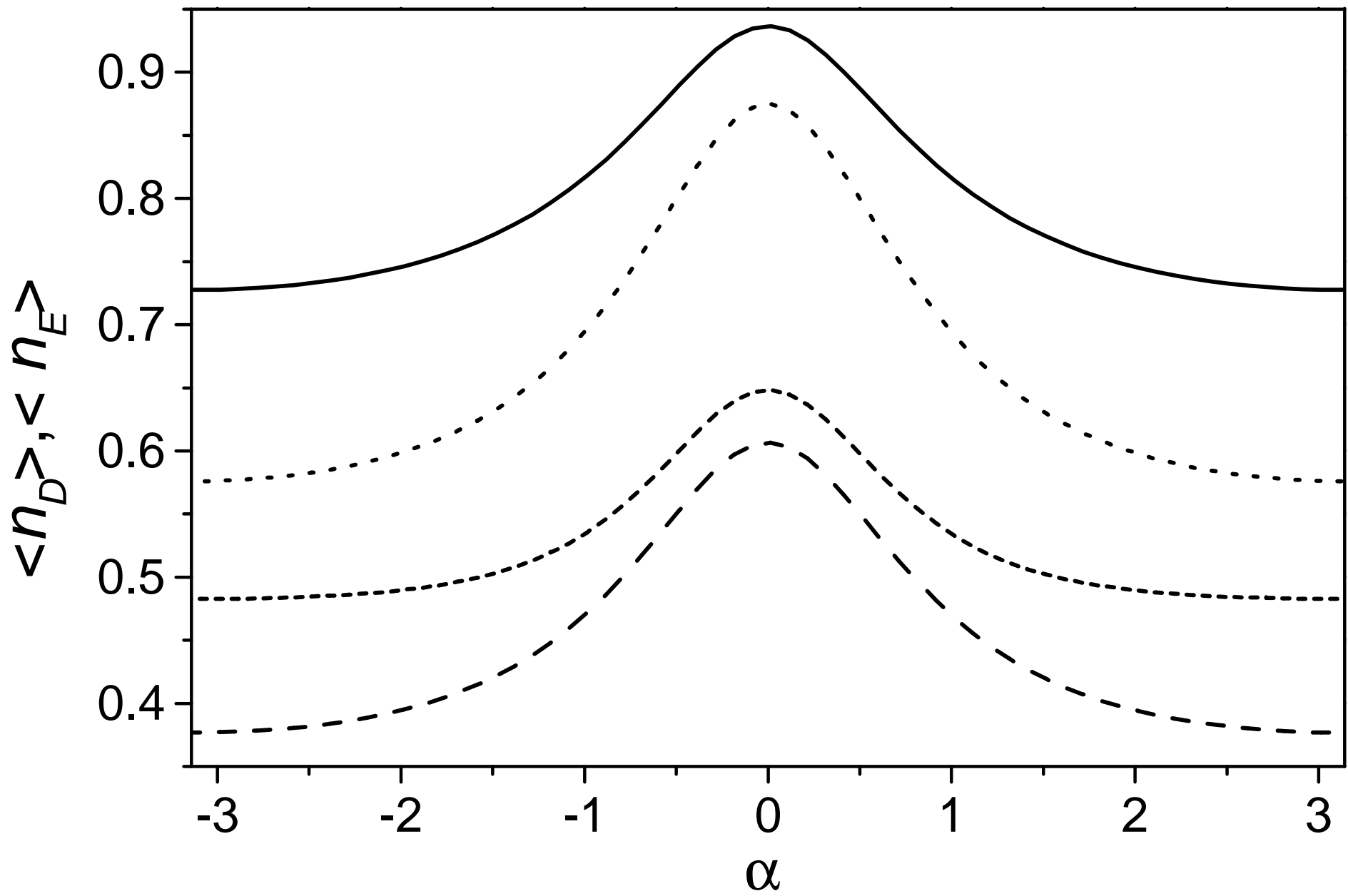
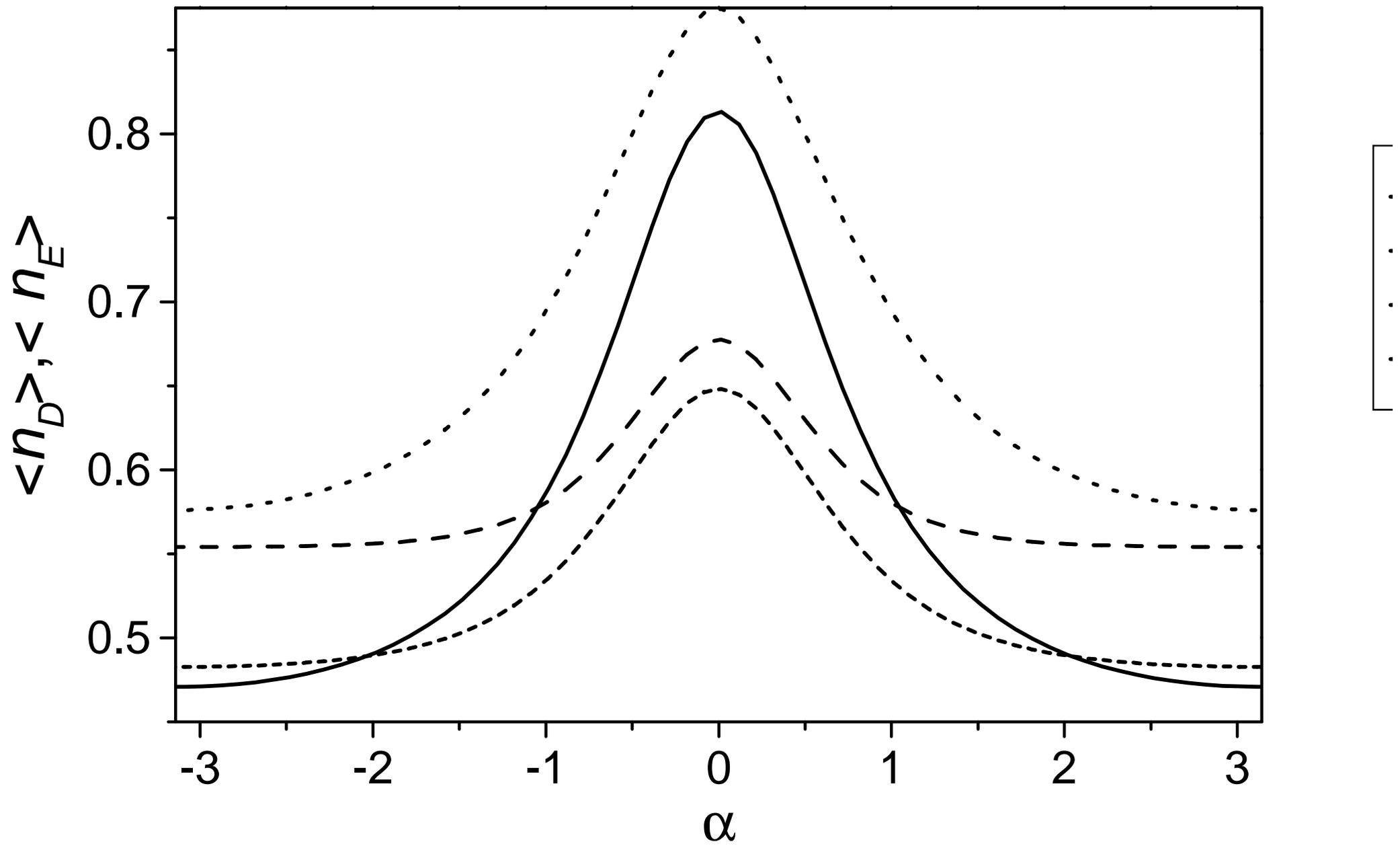
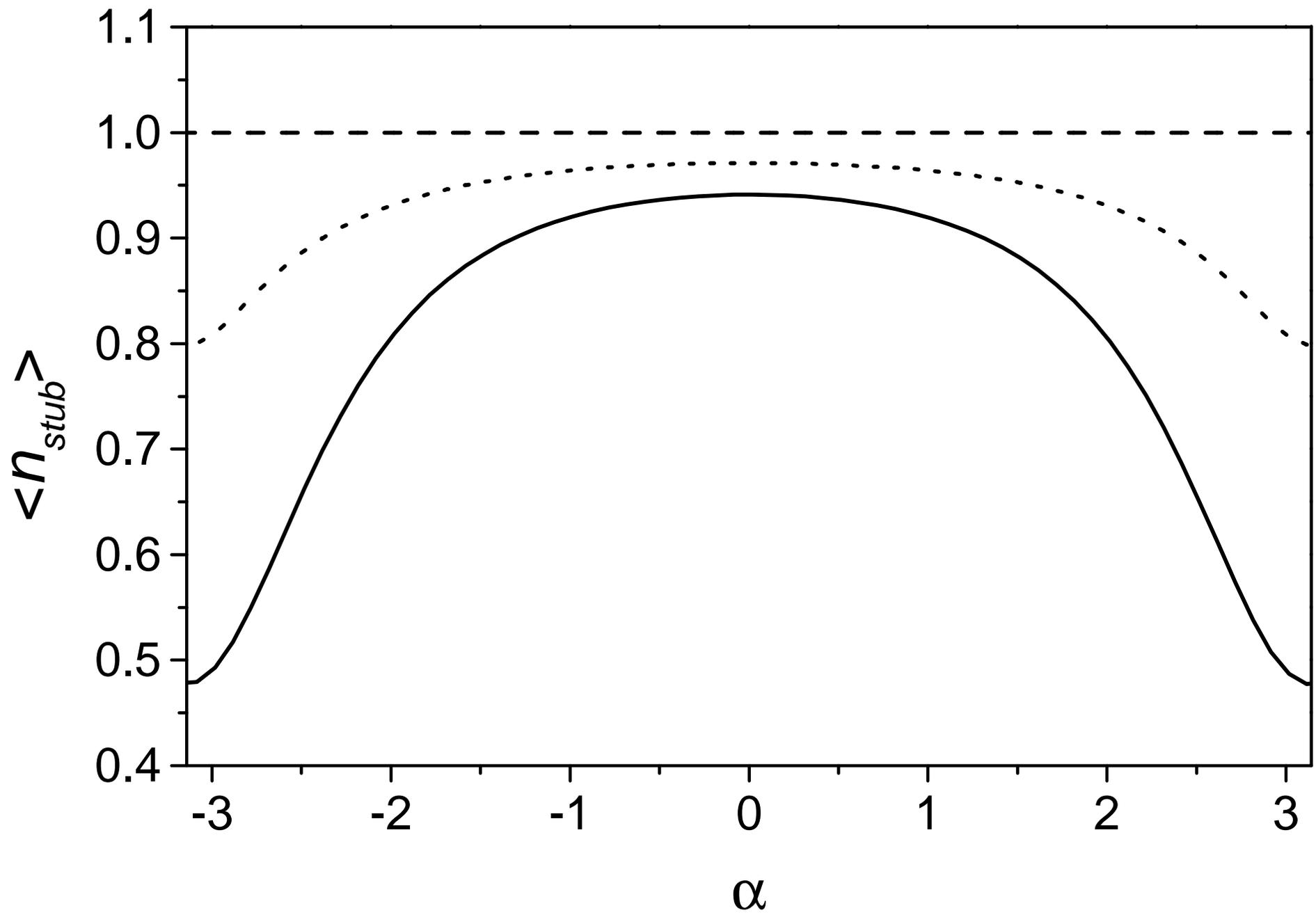
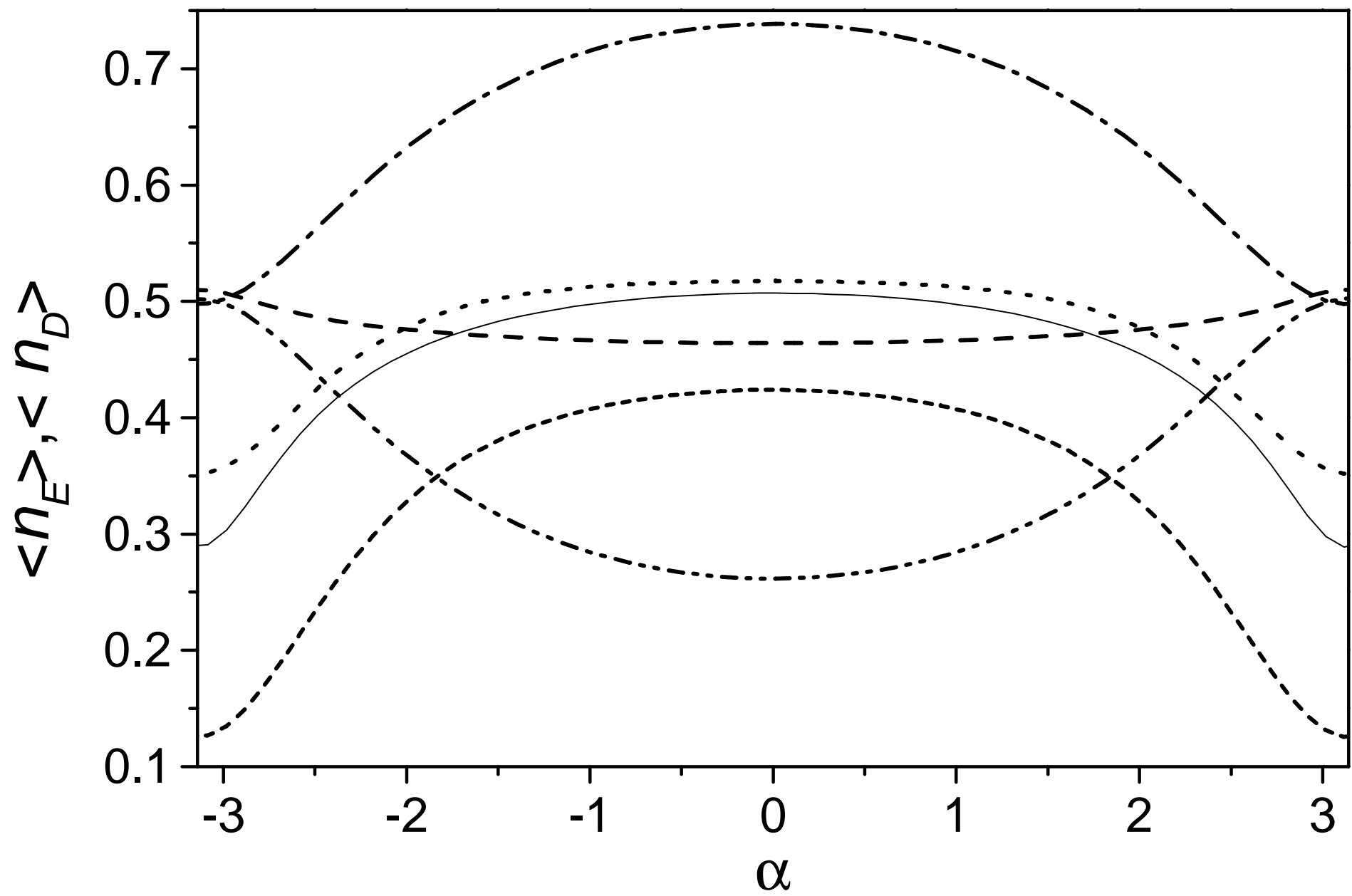
\end{document}